\documentclass[aps,prl,twocolumn]{revtex4}
\usepackage{amssymb}
\usepackage{graphicx}
\usepackage{subfigure}

\begin{document}

\title{A family of many-body models which are exactly solvable analytically}
\author{I. Fuentes-Schuller}
\thanks{Published before under maiden name Fuentes-Guridi}
\address{Instituto de Ciencias Nucleares, Universidad Nacional Aut\'onoma de M\'exico, A. Postal 70-543, M\'exico D.F. 04510, M\'exico}
\address{Perimeter Institute for Theoretical Physics, 31 Caroline St N, N2L 2Y5 Waterloo, Canada}
\author{P. Barberis-Blostein}
\address{Centro de Ciencias Fisicas, Universidad Nacional Aut\'onoma de M\'exico, Av. Universidad s/n, 
 Cuernavaca, M\'exico}
\begin{abstract}
We present a family of many-body models which are exactly solvable
analytically. The models are an extended
n-body interaction Lipkin-Meshkov-Glick model which considers spin-flip terms 
which are
associated with the interaction of an external classical field
which coherently manipulates the state of the system in order to,
for example, process quantum information. The models also describe a
two-mode Bose-Einstein condensate with a Josephson-type
interaction which includes n-particle elastic and inelastic
collisions. One of the models
corresponds to the canonical two-mode Bose-Einstein Hamitonian
plus a term which we argue must be considered in the
description of the two-mode condensate. Intriguingly, this extra term
allows for an exact and analytical solution of the two-particle collision two-mode BEC
problem. Our results open up an
arena to study many-body system properties analytically.
\end{abstract}

\maketitle
Many-body systems are of great relevance in most areas of physics.
In particular, there has been increasing interest in studying the properties of
many-body systems and learning how to manipulate them in order to
implement quantum information processing \cite{qc}. Ion traps, NMR systems,
optical lattices, spin chains, and many others have been
investigated for this purpose \cite{implementations}. Unfortunately, interesting many-body systems are rarely accessible
to purely analytic analysis. Exact solutions exist
mainly for one dimensional systems  and the higher dimensional
cases have to be treated numerically \cite{phase}. Numerical calculations are
in practice limited by the growing degrees of freedom of the system.
In nuclear physics, the Lipkin-Meshkov-Glick (LMG) model \cite{LMG} was
introduced as a toy model to study many-body properties. It has
been studied extensively because its integrability allows
for numerical analysis \cite{inte} and approximate solutions using the
algebraic Bethe ansatz \cite{bethe}. 
This model has relevance in quantum optics
since it is related to the two-mode Bose-Einstein condensate
(BEC) \cite{cirac}. In the study of BECs, multicomponent condensates are
of main interest. The lack of multimode BEC models which
have analytical solutions has restricted the understanding of
such mesoscopic systems whose most intriguing property is that they show collective quantum
behavior.

In this Letter we present a family of many-body models which are solvable analytically.
The family is characterize by the integer parameter $n$ which indicates de maximum number of particles
which interact in the system;
the n-model describes the n-body, (n-1)-body..., and 2-body interactions of N spin-1/2
particles which additionally interact with a classical coherent field.
The 2-model correspond to an extended LMG model which considers single spin-flip terms 
produced by the interaction with the
field and the effects of particle interactions during the spin-flip process. This
extension allows for an exact analytical solution of the model.
 
The n-model also describes a two-mode Bose-Einstein condensate
with a Josephson-type interaction. The canonical Josephson
Hamiltonian \cite{cirac} considers elastic two-particle collisions and has no
analytical solution. So far, the
canonical two-mode BEC model has not been
quantitatively verified by experiment although it qualitatively
describes some of the observed effects \cite{verification}. 

Our exactly soluble n-model corresponds to a
more complete two-mode BEC model since it considers all the features of the previously mentioned model, but
in addition considers all n-particle
elastic and inelastic collisions. Indeed, these multi-particle collisions must be
included in an realistic description of the condensate. It has been extensively pointed
out by experiment that inelastic
collisions \cite{inelastic} are present in the BEC and play an important role in
some systems. Moreover, higher order collisions are relevant beyond the dilute regime of BECs \cite{manycol,verification}.
So far only a couple of models incorporate
non-elastic collisions in multi-mode condensates \cite{luis}. 
Many-particle collisions have not been addressed at all in the theoretical models, although
it is known that they are physically relevant especially in the coldest phase
of the condensate where the particle density is high \cite{manycol}. 

Ironically, an effort is purposely made in the laboratory to suppress
many-particle collisions and inelastic processes, order to allow for comparison with
the existing theoretical models \cite{verification}. Here we show that including these processes in the theoretical description, 
the model becomes exactly soluble. 

We show, using the 2-model, that the
evolution of the relative population of the condensate presents collapse 
and revivals of Rabi oscillations.
We calculate the ground state of the system, and show that under certain circumstances, 
the ground state is in a multiple macroscopic superposition of coherent states.  

The analytical solutions of this family of models will allow for a deeper
understanding of many-body properties.

We introduce our models by considering the family of Hamiltonians 
$H^{n}_{0}=\sum_{i=0}^{n}A_{i}J_{z}^{i}$ where $J_{z}$ is some representation of the
SU(2) angular momentum operator in the $z$ direction and
${A_{i}}$ are real constants. Since
$J_{z}|j,m\rangle=m|j,m\rangle$ with $j$ and $m$ integers or half integers with
$m=-j,-j+1,...,j-1,j$, the eigenstates of the Hamiltonian are $|j,m\rangle$
with energy $\mathcal{E}^{n}_{m}=\sum_{i=0}^{n}A_{i}m^{i}$. By applying 
$U=e^{i\phi J_{z}}e^{i\theta J_{y}}$ which is the most general
rotation of $J_{z}$ in the SU(2) algebra with angles $\phi$ and
$\theta$ to $H_{0}^{n}$ we construct the family of n-models, 
\begin{eqnarray} \label{eq:model}
 H^{n}=U^{\dagger}H^{n}_{0}U=\sum_{i=0}^{n}A_{i}(U^{\dagger}J_{z}U)^{i}.
\end{eqnarray}
The exact and analytical solution of these Hamiltonians is
of course simply $U^{\dagger}|j,m\rangle$ with energy $\mathcal{E}_{m}^{n}$. The
integer parameter $n$ defines the n-model by considering up to $n$ powers of $J_{z}$ in
the $H^{n}_{0}$ Hamiltonian.
In terms of $J_{\pm}=J_{x}\pm iJ_{y}$, the 2-model
is
\begin{eqnarray}
 H_{2}&=&A_{1}(\cos{\theta}J_{z}+\sin{\theta}(e^{i\phi}J_{+}+e^{-i\phi}J_{-}))\\
&+&A_{2}(\cos^{2}{\theta}J^{2}_{z}+\sin^{2}{\theta}(e^{2i\phi}J^{2}_{+}+e^{-2i\phi}J^{2}_{-}\nonumber\\&+&J_{+}J_{-}+J_{-}J_{+})\nonumber\\
&+&\cos{\theta}\sin{\theta}(J_{z}(e^{i\phi}J_{+}+e^{-i\phi}J_{-})+h.c)).\nonumber
 \end{eqnarray}
The Hamiltonian can be written in terms of two bosonic operators $[a^{\dagger},a]=[b^{\dagger},b]=1$ 
through the Schwinger
representation which relates the bosonic opertaors to the angular momentum ones in
the following way: $J_{z}=a^{\dagger}a-b^{\dagger}b$,
$J_{+}=a^{\dagger}b$ and $J_{-}=ab^{\dagger}$. Note that 
the comutation relations of the SU(2) operators are indeed satisfied, 
and the total number operator $N=n_{a}+n_{b}=a^{\dagger}a+b^{\dagger}b$ is related to 
the total angular momentum by $J=N/2$. Therefore, by choosing different representations
of the SU(2) operators one can vary the total number of particles $N$. 
The 2-model in the two-mode representation is
{\ 
\begin{eqnarray}\label{eq:hbec}
H_{2}&=&A_{0}+\delta\omega(a^{\dagger}a-b^{\dagger}b)\\
&+&\lambda(e^{i\phi}a^{\dagger}b+e^{-i\phi}ab^{\dagger})+\mathcal{U}\, a^{\dagger}b^{\dagger}a b\nonumber\\
&+&\Lambda(e^{2i\phi}a^{\dagger}a^{\dagger}bb+h.c.)\nonumber\\
  &+&\mu((a^{\dagger}a^{\dagger}ab-b^{\dagger}a^{\dagger}ab)e^{i\phi}
  +h.c.),\nonumber
 \end{eqnarray}
with $A_{0}=A_{2}(\cos^{2}{\theta}N^{2}+\sin^{2}{\theta}N)$, $\delta\omega=A_{1}\cos{\theta}$, 
$\mathcal{U}=A_{2}(1-3\cos^{2}{\theta})$,
$\lambda=A_{1}\sin{\theta}$, $\mu=2\,A_{2}\cos{\theta}\sin{\theta}$ and
$\Lambda=A_{2}\sin^{2}{\theta}$. The states are 
related by $|j,m\rangle=|n_{a}=j+m\rangle\otimes|n_{b}=j-m\rangle$. Note that 
the particles are indistinguishable and the model only accounts for how many of them
are in a given state.
We will devote the rest of this paper to point out the physical
relevance of the family of Hamitonians given by Eq.(\ref{eq:model}).

The n-model describes the n-body, (n-1)-body,..., and 2-body interactions of 
$N=a^{\dagger}a+b^{\dagger}b$ spin-1/2 particles (with $n\leq N$) in the presence of a classical coherent field. 
We will first analyze the 2-model.
The first term in Eq.(\ref{eq:hbec}) describes the free energy of $a^{\dagger}a$ spin-1/2 particles in the spin 
up state and $b^{\dagger}b$ in the spin down 
state with frequency difference $\delta\omega$ . 
The interaction between two spins has strength $\mathcal{U}$ 
and corresponds to a dispersive process in which spins exchange their state while total spin is conserved. 
Additionally, we consider the interaction with an external classical field 
that produces one spin to flip state with coupling constant $\lambda$. 
The classical field could be an effective field due to the
presence of another system, other degrees of freedom of the
system or, more interesting, to an external experimenter
manipulating the state of the system using a laser. This last situation would be necessary for manipulation
quantum information in the system.

Due to the interaction of the field with the system, 
there is also a probability, parameterized by $\mu$, of having 
two spins flip their state. Since the
Hamiltonian has a second order character, i.e. it considerers
products of two and four creation and annihilation operators, one must consistently consider 
all possible second order physical processes. Thus, we include the two-particle spin-flip term ($\Lambda$)
and the term that describes a 
single dispersive process ($\mu$) taking place due to particle interaction while the laser produces 
a single spin to flip. 

Consider an experimenter using a laser to change the state of a single spin. In an
idealized situation the laser acts only on one spin and no 
interaction between spins is present.
But in a realistic situation the field cannot be directed 
only to one spin and there is a probability that two spins change their state. 
One must also consider the possibility having a single dispersive process while the laser acts on a single spin
due to spin interactions.
Therefore the terms corresponding to $\mu$ and $\Lambda$ 
model a more realistic situation.
It is remarkable that considering these extra terms allows for an analytical solution of the system.
 
Now we can analyze what the n-model describes: all possible n-body, (n-1)-body,...,and 2-body 
interactions with $n\leq N$. 
Thus, Eq. (\ref{eq:model}) describes the possibility of n spin-1/2 particles exchanging their state in such way
that the total spin is conserved and considers a laser which causes $m\leq n$ particles to flip state and 
all the possible dispersion terms accompanying this process.

If no classical field is applied, the model simply consists of N spin-1/2 particles which interact in such way
that the total spin is always conserved. The spin is conserved because no energy is provided to the system.

Surprisingly, ignoring the single spin-flip term ($\lambda$) 
and the term with a spin-flip plus single dispersion
($\mu$), the 2-model corresponds to
the LMG model of nuclear physics. The LMG model was constructed using products of two and four 
creation and annihilation operators 
with the purpose of creating a simple model for testing many-body
properties. It has so far no physical realization and it does not admit an exact analytical
solution. Here we showed that considering an extension to the model,
by considering consistently all possible products of two and four
creation and annihilation operators, an exact solution is found. 
Our extension also provides the model with a consistent physical picture.

Now lets focus on a closely related problem which does have a physical
realization. The family of Hamiltonians Eq.(\ref{eq:model}) is also of interest in
BEC since it describes
a two-mode BEC. Two mode BECs where first produced in JILA\cite{JILA2}
and MIT\cite{MIT}. The modes $a^{\dagger},a$ and
$b^{\dagger},b$ correspond to atoms in two different hyperfine
levels trapped by a magnetic potential with frequency difference
$\delta\omega$. The Hamiltonian considers the interaction with an
external laser to induce a Josephson-like coupling between the two
modes with coupling constant $\lambda$
and phase $\phi$ \cite{alternative}. This term corresponds to the
annihilation of one particle in one mode and the creation of one
particle in the other mode via the absorption or emission of a
laser photon. The terms with four bosonic operators describe 
two-particle elastic and spin-exchange inelastic collisions. The elastic
collisions have interaction
strength $\mathcal{U}$.
The inelastic collisions have interaction strength
$\mu$ when two particles of the same
specie collide and one of them is transformed into the other via
emission or absorption of a photon and
interaction strength $\Lambda$ when the collision transforms
two particles of one type into the other. 

Note that by fixing $\delta\omega$, $\lambda$ and
$U$ the inelastic collision constants, $\Lambda$ and $\mu$, are determined.
This is because in our model, the inelastic collisions are produced by the effect of
the laser on colliding particles. When the laser is applied to the
condensate to induce transitions between the hyperfine levels,
there is a probability of having a collision between two atoms and
a laser photon. Since the laser is considered to be a classical
field, this interaction gives rise to the two-particle inelastic
collision terms. This physical relationship is mathematically
expressed by the relationship between the coefficients. In our model the rate of elastic to inelastic collisions
is given by
\begin{equation} \label{eq:rate}
\frac{\mu+\Lambda}{\mathcal{U}}=\frac{\lambda}{2}\Big( \frac{\lambda+2\delta\omega}{A_{1}^{2}-3\delta\omega^{2}}\Big).
\end{equation}
Ignoring the inelastic terms in Eq.(\ref{eq:hbec}),
we find that our model
coincides with the canonical Josephson Hamiltonian \cite{cirac}
when the rate of collisions of same
species is equal. The assumption of equal 
collision rates for the same specie is also
made in \cite{cirac} order to find approximate and numerical solutions. 
Our model has the same number of free parameters as the canonical two-mode Hamiltonian. 
The only difference is that Hamiltonian in
Eq.(\ref{eq:hbec}) includes inelastic collisions, which are usually
present in real BECs \cite{inelastic}. In magnetic traps, inelastic collisions are commonly suppressed
to a large extent because they give rise to
atom losses. Spin exchange is the dominant loss mechanism in two-mode condensates \cite{stamper}.
But in optical traps the particle loss due to spin exchange is negligible and there it is no longer necessary to suppress 
the process \cite{stamper}.
Including the correct rate of inelastic collisions in the Hamiltonian allows for an 
analytical solution which is
simply $U^{\dagger}|N,m\rangle$. Note that $U$ in the Schwinger representation is the two-mode displacement operator 
$U=e^{\xi a^{\dagger}b+\xi^{\ast}ab^{\dagger}}$
where $\xi=\theta e^{i\phi}$. In \cite{berrybose} an author of this paper and
collaborators proposed the Hamiltonian $H_{2}$ to generate Berry phases in BECs but it was not
understood then that this could indeed correspond to a solution of the two-mode BEC.

We would like to emphasize that if for a specific physical system, the 
relation in Eq.(\ref{eq:rate}) between the rates of inelastic to elastic collisions does not hold or
cannot be arranged by external manipulation,
an analytical solution cannot be found using our method for such a condensate. 
Fortunately, in the laboratory
the rate of elastic and inelastic collisions can be 
manipulated by applying a magnetic potential \cite{roberts}
and it is possible to meet the experimental
values for the production ratios of those terms in a
two-mode BEC.

Let us now focus our attention on the 3-model Hamitonian $H_{3}$.
This Hamitonian corresponds to all possible three-particle and two-particle
interactions including elastic and
inelastic collisions between the same and different specie. The n-model in Eq.(\ref{eq:model}) describes
 a two-mode BEC where n-body, (n-1)-body,...,and 2-body interactions are considered. In
the two-mode BEC n-particle collision terms
are in principle present specially when the particle density is
high.

The canonical two-mode model \cite{cirac} considers
only two-particle elastic collisions and has no exact
analytical solution. Commonly, the Bethe anzatz is used to
find the ground and first excited state solution, or
numerical work is needed. The model introduced
here is more realistic, complete and has an exact analytical
solution. It is possible to analyze the whole spectrum and one need not
restrict the attention only to the ground state.

Due to the simplicity of our solution the ground state
$U^{\dagger}|N,m_0\rangle$ of $H_{2}$ is trivially found by
minimizing the energy $E^{2}_m=A_1 m+A_2 m^2$ with respect to m. For $A_2>0$, $m_0$ is the nearest integer to 
$-A_1/(2 A_2)$
or $m_0=-A_1 N/|A_1|$ when $|-A_1/(2 A_2)|>N$. For $A_2<0$, the minimum corresponds to $m_0=N$
if $A_1<0$ or $m_0=-N$ otherwise. 
The canonical two-mode BEC predicts that the ground state of
the condensate is, under certain conditions,
a macroscopic superposition of two coherent states \cite{cirac}. 
In Fig.~\ref{fig:m} we plot the relative population distribution for different
ground states of the second order of our model and find that macroscopic superpositions can involve 
several coherent states for $m_0\leq 1000$ as shown Fig.~\ref{fig:mc} and Fig.~\ref{fig:md}.
This difference must be due to inelastic collisions.
\begin{figure}[!ht]
\subfigure[$m_0=1000$]{
\includegraphics[width=1.5in]{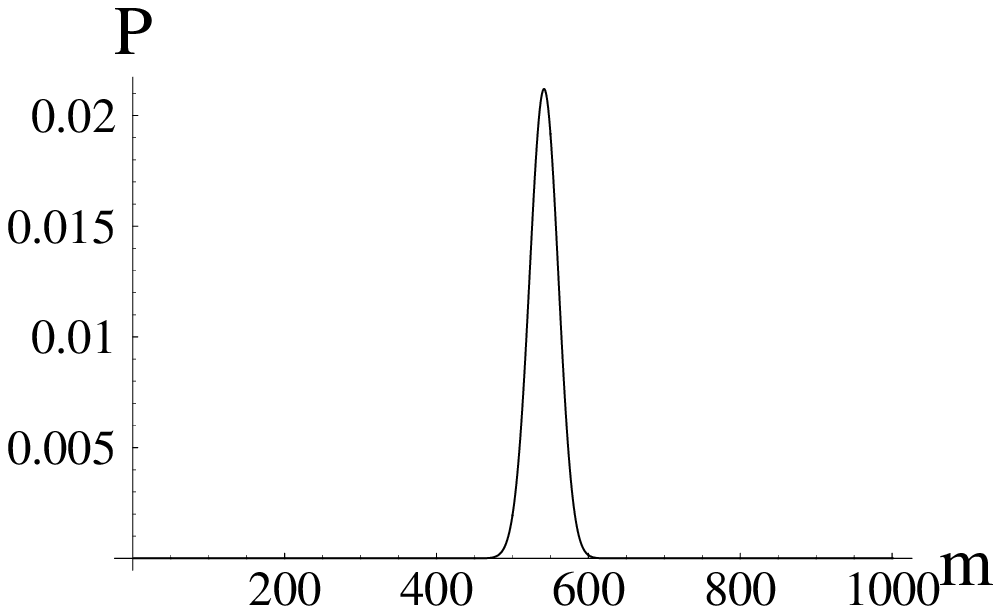}\label{fig:ma}}
\subfigure[$m_0=999$]{\includegraphics[width=1.5in]{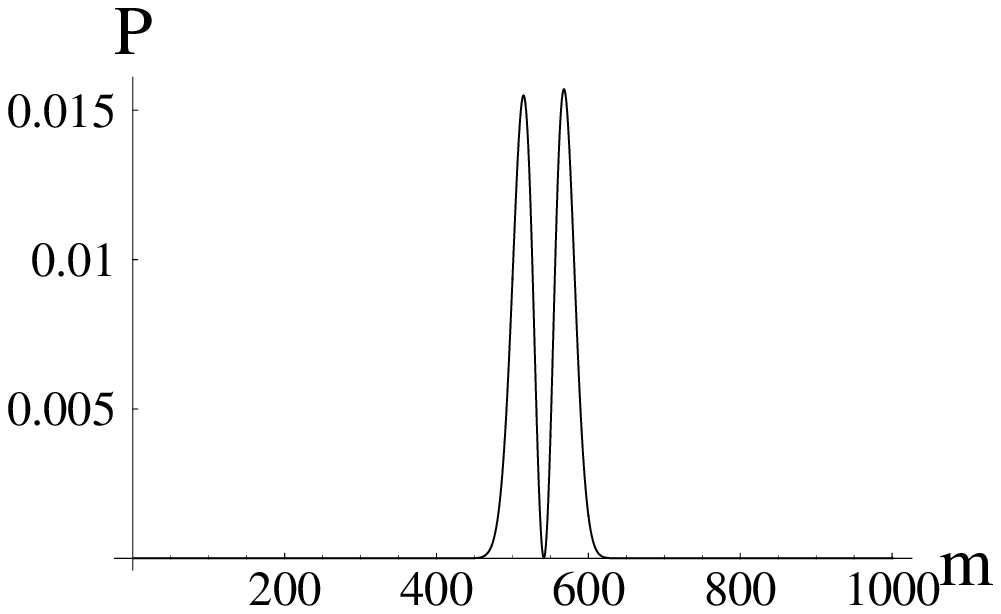}\label{fig:mb}}\\
\subfigure[$m_0=998$]{
\includegraphics[width=1.5in]{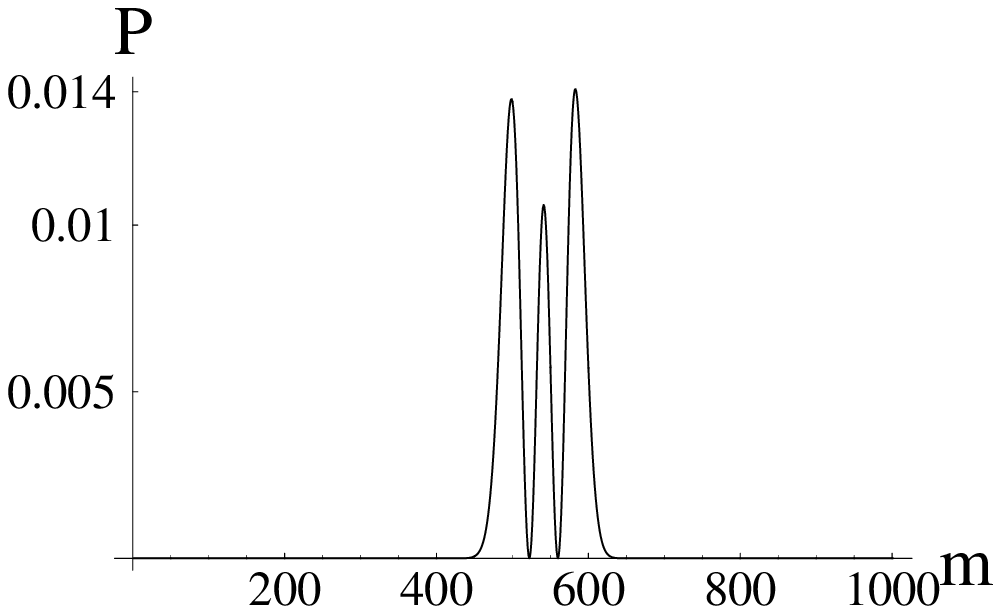}\label{fig:mc}}
\subfigure[$m_0=997$]{
\includegraphics[width=1.5in]{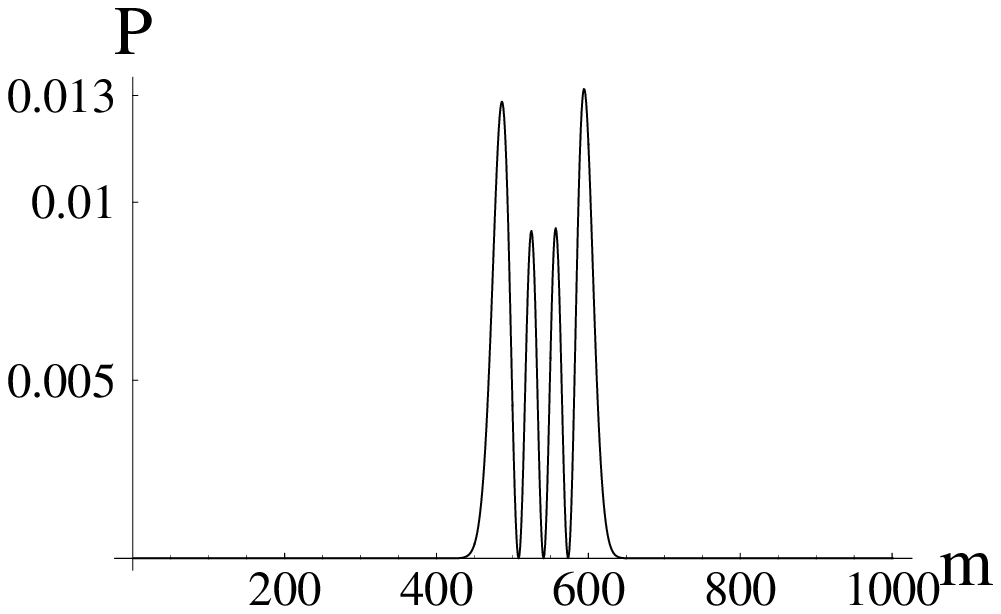}\label{fig:md}}
\caption{\label{fig:m} Ground state relative population distribution for $2001$ atoms. Different $m_0$
  correspond to different intensities of the laser. Quantum
  superposition appears when $m_0<1000$.}
\end{figure}

The evolution of the relative population $J_{z}$ for a given initial state $|\psi(t=0)\rangle=\sum_{m=-N}^{N}
C_m U^\dagger|N,m\rangle$, with coefficients $C_{m}$, is given by
\begin{eqnarray}
\langle j_z\rangle&=&-\sin\theta\sum_{-N+1}^{N} C_{m} C_{m-1} L_{m}\\ 
L_{m}&=&\cos(\phi+(E_{m-1}-E_m)\,t)(N(N+1)-m(m-1))\nonumber
\end{eqnarray}
The expectation value of $J_{y}$ which describes the evolution of the relative phase of the condensates
is equal to $J_{z}$ plus a phase shift of $\pi/2$. In Fig.~\ref{fig:jz} we plot the evolution of the
relative population for the initial state $|N,N\rangle$ where the condensate consists of a single specie.
The system presents Rabi-type oscillations with collapse and
revivals. We are currently studying the effects of higher-order collisions in the oscillations.
\begin{figure}[!ht]
\includegraphics[width=3in]{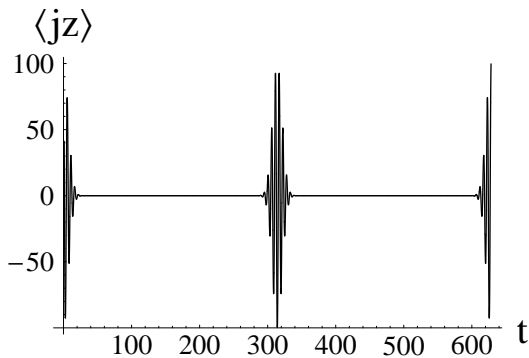}
\caption{\label{fig:jz} Evolution of $\langle j_z \rangle$. The initial
  condition is $|\psi(0)\rangle=|N=100,m=100\rangle$. $A_1=1$, $A_2=0.01$, $\theta=3/2$.}
\end{figure}
Interesting generalizations of our family of models which are currently under
study include a family of Hamiltonians with squeezing terms and models
replacing the SU(2) for SU(3) algebra. The first generalization is preformed by applying a
two-mode squeezing operator $S(\alpha)=e^{\alpha
a^{\dagger}b^{\dagger}-\alpha^{\ast} ab}$ to the Hamiltonian
$H_{0}^{n}$ which then has solution
$S^{\dagger}(\alpha)U^{\dagger}(\theta,\phi)|j,m\rangle$. The SU(3) model is found by
applying the most general rotation in the SU(3) algebra to a
polynomial in the SU(3) diagonal generators. The bosonic
representation of the SU(3) model describes a three mode (or
spin-1) BEC. In principle the model can be extended to the $SU(n)$
algebra corresponding to spin-J condensates. A general method of finding models
with exact analytical solution can be extrapolated from our model. For a given
algebra find the diagonal generators, construct a polynomial in them and
apply the most general rotation of the operators in the algebra to generate a new Hamiltonian.

The understanding of many-body systems in dimensions higher than 1 has been limited by
the lack of any realistic models with analytical solutions. The model we have introduced here allows for
the first time an analytical study of the n-body interactions of N spin-1/2 particles
in the presence of a classical coherent field and of a two-mode BEC
with n-body elastic and inelastic collisions.  
The model extends the Lipkin-Meshkov-Glick model of nuclear physics and the canonical two-mode BEC
models. The clear advantages of our model over these models include the possibility to study
higher order interactions.  
Currently we study inelastic and many-body interactions in BECs and their effects in the phase
transitions and entanglement properties of the system. We consider that the family of models
that we have introduced opens an arena to study higher-dimensional many-body systems analytically. 

We would like to thank O. Casta\~{n}os, J. Hirsh, O. Jimenez, J. Rogel-Salazar and P. Hess for interesting discussions.


\begin{thebibliography}
\bibitem{-}
\bibitem{qc} M. Nielsen and I. Chuang, Quantum Computation and Quantum Information, Cambridge University Press, 2000. 
\bibitem{implementations}Special issue on experimental proposals for quantum
computation, Fortschr. Phys. 48, No. 9–11 (2000).
\bibitem{phase} S. Sachdev, Quantum Phase Transitions, Cambridge
Univ. Press (1999); C. J. Thompson in Phase transitions and critical phenomena,
C. Domb and M. S. Green eds. (Academic Press,
London 1972).
\bibitem{LMG} H. J. Lipkin, N. Meshkov, and A. J. Glick, Nucl. Phys.
62, 188 (1965). 
\bibitem{inte}J. Dukelsky, S. Pittel, and G. Sierra, Rev. Mod. Phys.
76, 643 (2004).
 \bibitem{bethe} F. Pan and J. P. Draayer, Phys. Lett. B 451, 1 (1999); J. Links, H.-Q. Zhou, R. H. McKenzie, and M. D. Gould,
J. Phys. A 36, R63 (2003).
\bibitem{cirac}G. J. Milburn, J. Corney, E. M. Wright, and D. F. Walls,
Phys. Rev. A 55, 4318 (1997); J. I. Cirac, M. Lewenstein, K. Molmer, and P. Zoller,
Phys. Rev. A 57, 1208 (1998); A. J. Leggett, Rev. Mod. Phys. 73, 307–356 (2001).
\bibitem{verification} E. A. Cornell, J. R. Ensher and C. E. Wieman, "Experiments in Dilute Atomic Bose-Einstein Condensates", e-print cond-mat/9903109.
\bibitem{inelastic} J. Stenger, S. Inouye, D.M. Stamper-Kurn, H.-J. Miesner, A.P. Chikkatur, and W. Ketterle, Nature
396, 345 (1998).
\bibitem{manycol} Holzmann M., Krauth W., and Naraschewski M., e-print cond-mat/9806201.
\bibitem{luis} L. Santos, T. Pfau Spin-3 Chromium Bose-Einstein Condensates e-print cond-mat/0510634.
\bibitem{JILA2}C. J. Myatt et al., Phys. Rev. Lett. 78, 586 (1997).
 \bibitem{MIT} J. Stenger et al., Nature (London) 396, 345 (1998); H.-J.
Meisner et al., Phys. Rev. Lett. 82, 2228 (1999).
\bibitem{alternative} Alternatively, for the Na spinor
system of MIT, state dependent magnetic field gradient may be
applied to induce Josephson tunneling between two spatially
separated condensates.
\bibitem{stamper} D.M. Stamper-Kurn, et. al. Phys. Rev. Lett. 80, 0031-9007 (1997).
\bibitem{roberts} J.L. Roberts, N.R. Claussen, S.L. Cornish and C.E. Wieman, Phys. Rev. Lett. 85, 782 (2000).
\bibitem{berrybose} I. Fuentes-Guridi, J Pachos, S. Bose, V. Vedral, and S. Choi, Phys. Rev. A66, 022102 (2002). 
\end{thebibliography}
\end{document}